\documentstyle[twoside]{article}

\catcode`\@=11
\long\def\@makefntext#1{
\protect\noindent \hbox to 3.2pt {\hskip-.9pt  
$^{{\eightrm\@thefnmark}}$\hfil}#1\hfill}		

\def\thefootnote{\fnsymbol{footnote}}
\def\@makefnmark{\hbox to 0pt{$^{\@thefnmark}$\hss}}	
	
\def\ps@myheadings{\let\@mkboth\@gobbletwo
\def\@oddhead{\hbox{}
\rightmark\hfil\eightrm\thepage}   
\def\@oddfoot{}\def\@evenhead{\eightrm\thepage\hfil
\leftmark\hbox{}}\def\@evenfoot{}
\def\sectionmark##1{}\def\subsectionmark##1{}}

\oddsidemargin=\evensidemargin
\renewcommand{\thefootnote}{\fnsymbol{footnote}}

\newcounter{sectionc}\newcounter{subsectionc}\newcounter{subsubsectionc}
\renewcommand{\section}[1] {\vspace{12pt}\addtocounter{sectionc}{1} 
\setcounter{subsectionc}{0}\setcounter{subsubsectionc}{0}\noindent 
	{\tenbf\thesectionc. #1}\par\vspace{5pt}}
\renewcommand{\subsection}[1] {\vspace{12pt}\addtocounter{subsectionc}{1} 
	\setcounter{subsubsectionc}{0}\noindent 
	{\bf\thesectionc.\thesubsectionc. {\kern1pt \bfit #1}}\par\vspace{5pt}}
\renewcommand{\subsubsection}[1] {\vspace{12pt}\addtocounter{subsubsectionc}{1}
	\noindent{\tenrm\thesectionc.\thesubsectionc.\thesubsubsectionc.
	{\kern1pt \tenit #1}}\par\vspace{5pt}}
\newcommand{\nonumsection}[1] {\vspace{12pt}\noindent{\tenbf #1}
	\par\vspace{5pt}}

\newcounter{appendixc}
\newcounter{subappendixc}[appendixc]
\newcounter{subsubappendixc}[subappendixc]
\renewcommand{\thesubappendixc}{\Alph{appendixc}.\arabic{subappendixc}}
\renewcommand{\thesubsubappendixc}
	{\Alph{appendixc}.\arabic{subappendixc}.\arabic{subsubappendixc}}

\renewcommand{\appendix}[1] {\vspace{12pt}
        \refstepcounter{appendixc}
        \setcounter{figure}{0}
        \setcounter{table}{0}
        \setcounter{lemma}{0}
        \setcounter{theorem}{0}
        \setcounter{corollary}{0}
        \setcounter{definition}{0}
        \setcounter{equation}{0}
        \renewcommand{\thefigure}{\Alph{appendixc}.\arabic{figure}}
        \renewcommand{\thetable}{\Alph{appendixc}.\arabic{table}}
        \renewcommand{\theappendixc}{\Alph{appendixc}}
        \renewcommand{\thelemma}{\Alph{appendixc}.\arabic{lemma}}
        \renewcommand{\thetheorem}{\Alph{appendixc}.\arabic{theorem}}
        \renewcommand{\thedefinition}{\Alph{appendixc}.\arabic{definition}}
        \renewcommand{\thecorollary}{\Alph{appendixc}.\arabic{corollary}}
        \renewcommand{\theequation}{\Alph{appendixc}.\arabic{equation}}
        \noindent{\tenbf Appendix \theappendixc #1}\par\vspace{5pt}}
\newcommand{\subappendix}[1] {\vspace{12pt}
        \refstepcounter{subappendixc}
        \noindent{\bf Appendix \thesubappendixc. {\kern1pt \bfit #1}}
	\par\vspace{5pt}}
\newcommand{\subsubappendix}[1] {\vspace{12pt}
        \refstepcounter{subsubappendixc}
        \noindent{\rm Appendix \thesubsubappendixc. {\kern1pt \tenit #1}}
	\par\vspace{5pt}}

\newcommand{\textlineskip}{\baselineskip=13pt}
\newcommand{\smalllineskip}{\baselineskip=10pt}

\def\eightcirc{
\begin{picture}(0,0)
\put(4.4,1.8){\circle{6.5}}
\end{picture}}
\def\eightcopyright{\eightcirc\kern2.7pt\hbox{\eightrm c}} 

\newcommand{\copyrightheading}[1]
	{\vspace*{-2.5cm}\smalllineskip{\flushleft
	{\footnotesize Modern Physics Letters A, #1}\\
	{\footnotesize $\eightcopyright$\, World Scientific Publishing
	 Company}\\
	 }}

\newcommand{\publisher}[2]{{\begin{center}\footnotesize\smalllineskip 
	Received #1\\
	Revised #2
	\end{center}
	}}

\def\abstracts#1#2#3{{
	\centering{\begin{minipage}{4.5in}\baselineskip=10pt\footnotesize
	\parindent=0pt #1\par 
	\parindent=15pt #2\par
	\parindent=15pt #3
	\end{minipage}}\par}} 

\def\keywords#1{{
	\centering{\begin{minipage}{4.5in}\baselineskip=10pt\footnotesize
	{\footnotesize\it Keywords}\/: #1
	 \end{minipage}}\par}}

\newcommand{\bibit}{\nineit}
\newcommand{\bibbf}{\ninebf}
\renewenvironment{thebibliography}[1]
	{\frenchspacing
	 \ninerm\baselineskip=11pt
	 \begin{list}{\arabic{enumi}.}
        {\usecounter{enumi}\setlength{\parsep}{0pt}     
	 \setlength{\leftmargin 12.7pt}{\rightmargin 0pt} 
         \setlength{\itemsep}{0pt} \settowidth
	{\labelwidth}{#1.}\sloppy}}{\end{list}}

\newcounter{itemlistc}
\newcounter{romanlistc}
\newcounter{alphlistc}
\newcounter{arabiclistc}

\newcommand{\fcaption}[1]{
        \refstepcounter{figure}
        \setbox\@tempboxa = \hbox{\footnotesize Fig.~\thefigure. #1}
        \ifdim \wd\@tempboxa > 5in
           {\begin{center}
        \parbox{5in}{\footnotesize\smalllineskip Fig.~\thefigure. #1}
            \end{center}}
        \else
             {\begin{center}
             {\footnotesize Fig.~\thefigure. #1}
              \end{center}}
        \fi}

\newcommand{\tcaption}[1]{
        \refstepcounter{table}
        \setbox\@tempboxa = \hbox{\footnotesize Table~\thetable. #1}
        \ifdim \wd\@tempboxa > 5in
           {\begin{center}
        \parbox{5in}{\footnotesize\smalllineskip Table~\thetable. #1}
            \end{center}}
        \else
             {\begin{center}
             {\footnotesize Table~\thetable. #1}
              \end{center}}
        \fi}

\def\@citex[#1]#2{\if@filesw\immediate\write\@auxout
	{\string\citation{#2}}\fi
\def\@citea{}\@cite{\@for\@citeb:=#2\do
	{\@citea\def\@citea{,}\@ifundefined
	{b@\@citeb}{{\bf ?}\@warning
	{Citation `\@citeb' on page \thepage \space undefined}}
	{\csname b@\@citeb\endcsname}}}{#1}}

\newif\if@cghi
\def\cite{\@cghitrue\@ifnextchar [{\@tempswatrue
	\@citex}{\@tempswafalse\@citex[]}}
\def\citelow{\@cghifalse\@ifnextchar [{\@tempswatrue
	\@citex}{\@tempswafalse\@citex[]}}
\def\@cite#1#2{{$\null^{#1}$\if@tempswa\typeout
	{IJCGA warning: optional citation argument 
	ignored: `#2'} \fi}}

\def\pmb#1{\setbox0=\hbox{#1}
	\kern-.025em\copy0\kern-\wd0
	\kern.05em\copy0\kern-\wd0
	\kern-.025em\raise.0433em\box0}

\def\fnt#1#2{\footnotetext{\kern-.3em
	{$^{\mbox{\scriptsize #1}}$}{#2}}}

\def\fpage#1{\begingroup
\voffset=.3in
\thispagestyle{empty}\begin{table}[b]\centerline{\footnotesize #1}
	\end{table}\endgroup}

\def\runninghead#1#2{\pagestyle{myheadings}
\markboth{{\protect\footnotesize\it{\quad #1}}\hfill}
{\hfill{\protect\footnotesize\it{#2\quad}}}}
\font\tenrm=cmr10
\font\tenit=cmti10 
\font\tenbf=cmbx10
\font\bfit=cmbxti10 at 10pt
\font\ninerm=cmr9
\font\nineit=cmti9
\font\ninebf=cmbx9
\font\eightrm=cmr8

\def\qed{\hbox{${\vcenter{\vbox{			
   \hrule height 0.4pt\hbox{\vrule width 0.4pt height 6pt
   \kern5pt\vrule width 0.4pt}\hrule height 0.4pt}}}$}}

\renewcommand{\thefootnote}{\fnsymbol{footnote}}	

\include{epsf}
\begin{document}

\runninghead{I. Stancu}{Can the Super-Kamiokande Atmospheric Data Predict the
                        Solar Neutrino Deficit ?}

\normalsize\textlineskip
\thispagestyle{empty}
\setcounter{page}{1}

\copyrightheading
                  {UCRHEP-T250}

\vspace*{0.88truein}

\fpage{1}
\centerline{\bf CAN THE SUPER-KAMIOKANDE ATMOSPHERIC DATA}
\vspace*{0.035truein}
\centerline{\bf PREDICT THE SOLAR NEUTRINO DEFICIT ?}
\vspace*{0.37truein}
\centerline{\footnotesize ION STANCU\footnote{E-mail: ion.stancu@ucr.edu}}
\vspace*{0.015truein}
\centerline{\footnotesize \it Department of Physics, University of California}
\baselineskip=10pt
\centerline{\footnotesize \it Riverside, CA 92521, USA}
\vspace*{0.225truein}
\publisher{29 March 1999}{31 March 1999}

\vspace*{0.21truein}
\abstracts{
In this Letter we show that the evidence for neutrino oscillations from the
Super-Kamiokande atmospheric neutrino data fully determines the $3 \times 3$
neutrino-oscillations mixing matrix and predicts an energy independent solar
neutrino deficit at the level of 45\%.
This corresponds to a ratio of measured to predicted neutrino flux of
$R_e^{Solar} = 0.55$, in good agreement with the experimental results.
We achieve this result within the framework of a minimal, three-generations
neutrino mixing, with mass squared differences of
$\Delta M^2 \simeq 0.45$~eV$^2$
and
$\Delta m^2 = {\cal O}(10^{-3})$~eV$^2$.
The mixing matrix derived here is characterized by the mixing angles
$\theta = 35.1^\circ$, $\beta = 5.5^\circ$, and $\psi = 23.3^\circ$, and
a vanishing CP-violating phase, $\delta = 0$.
}{}{}

\vspace*{10pt}

\def\beq{\begin{eqnarray}}
\def\eeq{\end  {eqnarray}}


\vspace*{1pt}\textlineskip	
\section{Introduction}		
\vspace*{-0.5pt}
\noindent
The phenomenon of neutrino oscillations was first postulated by
Pontecorvo\cite{Pontecorvo} in 1957, and has been ever since a prime candidate
in explaining the long-standing puzzle provided by the solar neutrino
deficit.\cite{Homestake,Kamiokande,Sage,Gallex,SkSol}
The anomalies reported in the atmospheric neutrino
production\cite{Hirata,Becker,SkAtm} and the positive signals obtained by LSND
in both the $\bar\nu_\mu \to \bar\nu_e$ and the $\nu_\mu \to \nu_e$
appearance channels\cite{LsndDar935,LsndDif935} have contributed to
further strengthen the evidence towards neutrino oscillations.
At the same time, these observations raise the question whether they can all
be explained in the minimal, three-generation neutrino mixing formalism.
Typically, all experiments have been analyzed in terms of the simpler,
two-neutrino oscillation hypothesis, which in turn have given rise to three
different mass squared differences, $\delta m^2$.
The solar neutrino deficit implies a $\delta m^2$ of
${\cal O}(10^{-10})$~eV$^2$ or
${\cal O}(10^{-5})$~eV$^2$, corresponding to vacuum or MSW oscillations,
respectively.
The atmospheric neutrino deficit indicates a $\delta m^2$ of
${\cal O}(10^{-3})$~eV$^2$.
The LSND evidence points towards a $\delta m^2$ of ${\cal O}(1)$~eV$^2$.
Since only two mass squared differences are independent, one possible solution
to this puzzle is the introduction of a fourth neutrino flavor, which must be
sterile if its mass is low.
Another solution is to ignore one of the indications listed above, typically
LSND, as it is the only evidence that has not been independently confirmed by
another experiment.
(However, one should keep in mind that the LSND evidence comes from two
different channels, with different neutrino fluxes, signatures, backgrounds and
systematics.)
On the other hand, if one assumes that the LSND excess is indeed due to
neutrino oscillations, the two mass squared differences implied by it and by
the Super-Kamiokande atmospheric data would necessarily lead to an energy
independent oscillation probability for the solar neutrinos.

The latter approach has been followed by several
authors.\cite{Thun,Barenboim,Teshima}
Using three data points,
(1) a flat solar neutrino ratio of $R_e^{Solar} = 0.5$,
(2) the atmospheric $\nu_\mu$ deficit for up-going muons
    (interpreted as $\nu_\mu \to \nu_\tau$ oscillations), and
(3) the oscillation probability reported by LSND,
they fully determine the neutrino-mixing matrix.
This, together with the LSND and Super-Kamiokande mass squared differences,
is shown to be consistent with all current neutrino oscillations experiments.
Alternatively, a more elaborate fit to all data,\cite{Ohlsson} yields similar
results.
What we show in this Letter is that the Super-Kamiokande atmospheric data alone
contains enough information, through its explicit $L/E$ dependence, to fully
determine the neutrino-oscillations mixing matrix, and is able to predict the
solar neutrino deficit to a level which is consistent will the current
measurements.

\textheight=7.8truein
\setcounter{footnote}{0}
\renewcommand{\thefootnote}{\alph{footnote}}

\section{Formalism}
\noindent
Throughout this paper we shall assume that the three neutrino weak eigenstates
$(\nu_e, \nu_\mu, \nu_\tau)$ are linear superpositions of three mass
eigenstates $(\nu_1, \nu_2, \nu_3)$,
\beq
|\nu_\ell> = \sum_{j=1}^3 U_{\ell j} |\nu_j> \hspace{10mm} (\ell=e,\mu,\tau),
\eeq
with masses $m_1 < m_2 < m_3$.
The unitary matrix $U$, parameterized in terms of three mixing angles
$(\theta,\beta,\psi)$ and a CP-violating phase ($\delta$), reads
\beq
U = \left(
\begin{array}{ccc}
  c_\theta c_\beta                                       &
  s_\theta c_\beta                                       &
  s_\beta                                                \\
- c_\theta s_\beta s_\psi e^{i\delta} - s_\theta c_\psi  &
  c_\theta c_\psi - e^{i\delta} s_\theta s_\beta s_\psi  &
  c_\beta s_\psi e^{i\delta}                             \\
- c_\theta s_\beta c_\psi + s_\theta s_\psi e^{-i\delta} &
- s_\theta s_\beta c_\psi - c_\theta s_\psi e^{-i\delta} &
  c_\beta c_\psi
\end{array}
\right)
\eeq
in the Maiani representation,\cite{Maiani} where $c_\theta=\cos\theta$,
$s_\theta=\sin\theta$, etc.
Consequently, assuming that the mass eigenstates are relativistic\cite{DvaTg}
and stable, the oscillation probability from a state $|\nu_\ell>$ to a state
$|\nu_{\ell'}>$ is given by
\beq
P(\nu_\ell \to \nu_{\ell'}) = \delta_{\ell\ell'} -
\sum_{i=1}^{2} \sum_{j=i+1}^{3} & \!\!\!\!\!\!\!\!\!\! &
\left[ 4 \, Re(U_{\ell'i}U_{\ell i}^*U_{\ell'j}^*U_{\ell j})
\sin^2 \left( 1.27 \, \Delta m_{ij}^2 \frac{L}{E} \right) \right.
\nonumber\\
& \!\!\!\!\!\!\!\!\!\! & \left.
\!\! - 2 \, Im(U_{\ell'i}U_{\ell i}^*U_{\ell'j}^*U_{\ell j})
\sin \, \left( 2.54 \, \Delta m_{ij}^2 \frac{L}{E} \right) \right].
\label{Posc}
\eeq
In Eq.(\ref{Posc}) above $E$ is the neutrino energy (expressed in MeV), $L$ is
the distance between the generation point and the detection point (expressed in
m), and $\Delta m_{ij}^2 \equiv m_i^2-m_j^2$ (expressed in eV$^2$).
The neutrino mass spectrum is characterized by two mass squared differences,
which we shall loosely refer to as ``mass scales'' henceforth, $\Delta m^2$
and $\Delta M^2$:
\beq
\Delta m^2 & \!\!\!\! \equiv \!\!\!\! & \Delta m_{21}^2 = m_2^2 - m_1^2,\\
\Delta M^2 & \!\!\!\! \equiv \!\!\!\! & \Delta m_{31}^2 = m_3^2 - m_1^2.
\eeq
The $\Delta m^2$ mass scale is taken to be of ${\cal O}(10^{-3})$~eV$^2$, as
suggested by the Super-Kamiokande atmospheric data, while the $\Delta M^2$ mass
scale is taken to be of ${\cal O}(1)$~eV$^2$, as suggested by LSND.
Furthermore, we shall only consider the case in which there is no CP-violation
in the neutrino sector, and thus we shall set the phase $\delta = 0$.

\section{The Super-Kamiokande Atmospheric Neutrino Data}
\noindent
Let us assume that at the top of the atmosphere, at $t = 0$, the number of
muon and electron neutrinos produced are $N_\mu$ and $N_e$, respectively.
Although both neutrinos and antineutrinos are produced in both flavors, we
shall use the terms ``muon neutrinos'' and ``electron neutrinos'' loosely, to
include both neutrinos and antineutrinos.
The ratio of muon to electron neutrinos, $r$,
\beq
r = \frac{N(\nu_\mu + \bar\nu_\mu)}{N(\nu_e + \bar\nu_e)} = \frac{N_\mu}{N_e}
\eeq
is, in general, a monotonically increasing function of energy.
However, for the relevant range of $L/E$ values in Super-Kamiokande, $r$ may
be assumed constant $(r = 2)$, as we shall do throughout this Letter.
Within a detector at a distance $L \simeq t$ from the production point, the
number of muon and electron neutrinos, $N_\mu'$ and $N_e'$, respectively,
are given by:
\beq
N_\mu' & \!\!\!\! = \!\!\!\! & N_\mu P_{\mu \mu} + N_e   P_{e \mu},\\
N_e'   & \!\!\!\! = \!\!\!\! & N_e   P_{e   e  } + N_\mu P_{\mu e},
\eeq
where $P_{\mu \mu}$ is the $\nu_\mu$ survival probability:
\beq
P_{\mu \mu} = 1 & \!\!\!\! - \!\!\!\! &
                      4 \, U_{\mu 1}^2U_{\mu 2}^2 \, \sin^2
                      \left( 1.27 \, \Delta m^2 \, \frac{L}{E} \right)
                    - 4 \, U_{\mu 1}^2U_{\mu 3}^2 \, \sin^2
                      \left( 1.27 \, \Delta M^2 \, \frac{L}{E} \right)
\nonumber\\
                & \!\!\!\! - \!\!\!\! &
                      4 \, U_{\mu 2}^2U_{\mu 3}^2 \, \sin^2
                      \left[ 1.27 \, (\Delta M^2 - \Delta m^2) \, \frac{L}{E}
                      \right],
\label{Pmumu}
\eeq
$P_{e \mu}$ is the neutrino oscillation probability $P(\nu_e \to \nu_\mu)$:
\beq
P_{e \mu}  = & \!\!\!\! - \!\!\!\! &
                   4 \, U_{e1}U_{e2}U_{\mu1}U_{\mu2} \, \sin^2
                   \left( 1.27 \, \Delta m^2 \, \frac{L}{E} \right)
\nonumber\\
             & \!\!\!\! - \!\!\!\! &
                   4 \, U_{e1}U_{e3}U_{\mu1}U_{\mu3} \, \sin^2
                   \left( 1.27 \, \Delta M^2 \, \frac{L}{E} \right)
\nonumber\\
             & \!\!\!\! - \!\!\!\! &
                   4 \, U_{e2}U_{e3}U_{\mu2}U_{\mu3} \, \sin^2
                   \left[ 1.27 \, (\Delta M^2 - \Delta m^2) \, \frac{L}{E}
                   \right],
\label{Pemu}
\eeq
and $P_{e e}$ is the $\nu_e$ survival probability:
\beq
P_{ee} = 1 & \!\!\!\! - \!\!\!\! &
                 4 \, U_{e1}^2U_{e2}^2 \, \sin^2
                 \left( 1.27 \, \Delta m^2 \, \frac{L}{E} \right)
               - 4 \, U_{e1}^2U_{e3}^2 \, \sin^2
                 \left( 1.27 \, \Delta M^2 \, \frac{L}{E} \right)
\nonumber\\
           & \!\!\!\! - \!\!\!\! &
                 4 \, U_{e2}^2U_{e3}^2 \, \sin^2
                 \left[ 1.27 \, (\Delta M^2 - \Delta m^2) \, \frac{L}{E}
                 \right].
\label{Pee}
\eeq
Since we are only considering the scenario with a vanishing CP-violating phase
$\delta = 0$, $P_{\mu e} = P_{e \mu}$.
The ratios of measured to predicted (without neutrino oscillations) $\mu$-like
and $e$-like events at Super-Kamiokande, $R_{\mu}$ and $R_e$, respectively,
are
\beq
R_\mu & \!\!\!\! = \!\!\!\! & \frac{N_\mu'}{N_\mu}
                 = P_{\mu \mu} + 0.5 P_{e \mu},
\label{RmuDef}\\
R_e   & \!\!\!\! = \!\!\!\! & \frac{N_e'  }{N_e  }
                 = P_{e   e  } + 2.0 P_{\mu e},
\label{ReDef}
\eeq
where we have explicitly set $r = 2$.
It is precisely these two ratios, and in particular their $L/E$ dependence - as
measured by the Super-Kamiokande group - that we shall exploit in order to
fully determine the neutrino oscillations mixing matrix.

For the mass scales considered here, something remarkable happens in the 3rd
and in the 8th bin of the Super-Kamiokande atmospheric neutrino data -- see
Fig.4 in Ref.[9].
In the 3rd bin, corresponding to
$10^{1.5} \,\, \mbox{m/MeV} < L/E < 10^2 \,\, \mbox{m/MeV}$,
the oscillations involving the $\Delta m^2$ mass scale are approximately zero,
and thus negligible, whereas the oscillations involving the $\Delta M^2$ mass
scale average out to 1/2.
One can certainly argue about how many oscillations it would take in a given
$L/E$ interval in order to safely assume that the oscillations do indeed
average out.
For instance, for $\Delta M^2 = 0.18$~eV$^2$ one has approximately five
oscillations in this interval, whereas $\Delta M^2 = 0.36$~eV$^2$ yields
obviously twice as many -- approximately ten oscillations.
In our view, even these relatively low numbers of oscillations do allow to
effectively average out the sine squared variation.
Therefore, we shall henceforth impose a weak lower bound of about
$0.27$~eV$^2$ for the $\Delta M^2$ mass scale, as the average of the two values
mentioned above.
As we shall see later on, the value of $\Delta M^2$ favored by this analysis
lies well above this limit.
In the 8th bin, corresponding to
$10^4 \,\, \mbox{m/MeV} < L/E < 10^{4.5} \,\, \mbox{m/MeV}$,
all oscillations average out to 1/2.
(For example, for $\Delta m^2 = 10^{-3}$~eV$^2$ one has approximately nine
oscillations in this $L/E$ interval, and thus our assumption is well
justified.)

\begin{figure}[htb]
\vspace*{13pt}
\begin{center}
\mbox{
\epsfxsize=8cm \epsfbox[20 35 550 550]{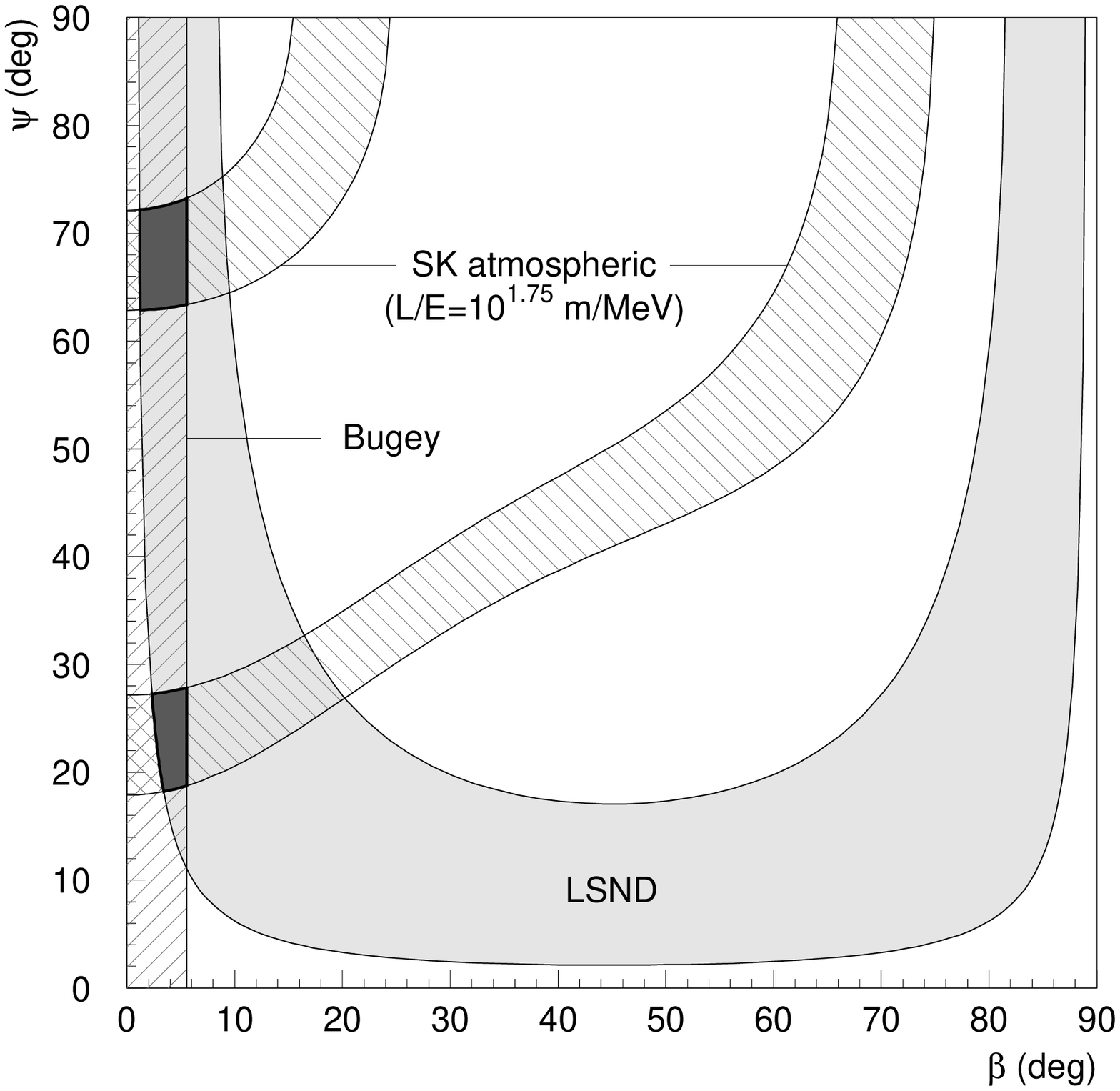}}
\end{center}
\vspace*{13pt}
\fcaption{
Allowed regions for the mixing angles $\beta$ and $\psi$ from the
Super-Kamiokande ratio of ratios in the 3rd $L/E$ bin (see text), from the
the LSND {\footnotesize \it appearance} signal (grey-shaded area), and from the
Bugey {\footnotesize \it disappearance} measurement (hatched area left of
$\beta = 5.5^\circ$).
}
\label{Fig-R175}
\end{figure}

Following these arguments, the ratios of measured to expected $\mu$-like and
$e$-like events, as defined by Eqs.(\ref{RmuDef})--(\ref{ReDef}), are
independent of the underlying mass scales in these two bins.
For the 3rd bin, these ratios yield a rather simple form:
\beq
R_\mu^{(3)} & \!\!\!\! = \!\!\!\! & 1 - c^2_\beta s^2_\psi \,
                    (2 - s^2_\beta - 2 c^2_\beta s^2_\psi),\\
R_e^{(3)}   & \!\!\!\! = \!\!\!\! & 1 - 2 c^2_\beta s^2_\beta \,
                                   (1 - 2 s^2_\psi).
\eeq
They depend only on the 1-3 and 2-3 mixing angles, $\beta$ and $\psi$,
respectively.
One could argue that using the values measured by Super-Kamiokande,
$R_\mu^{(3)} = 0.908 \pm 0.068$ and $R_e^{(3)} = 1.225 \pm 0.108$, directly
determine these two angles.
However, at this point we prefer to only restrict the possible values of
$\beta$ and $\psi$ from the ratio of ratios, $R_\mu^{(3)}/R_e^{(3)} = 0.741
\pm 0.088$.
This is a much more robust quantity, as the large systematic errors in the
individual flux calculations largely cancel out.
The allowed values for the mixing angles $\beta$ and $\psi$ are shown in
Fig.~\ref{Fig-R175}, at the $\pm 1 \sigma$ level.
The expressions for the $\mu$-like and $e$-like event ratios in the 8th bin,
$R_\mu^{(8)}$ and $R_e^{(8)}$, respectively, are rather long and cumbersome,
and therefore we choose not to display them explicitly.
With four data points (the ratios of ratios and the $\mu$-like ratios at the
two bins) one can fit the three mixing angles, along with an overall flux
normalization constant.
This procedure determines the mixing angles to be $\theta = 8.3^\circ$,
$\beta = 52.6^\circ$, and $\psi = 49.8^\circ$, for an overall scale factor of
0.88.
The ratio of the measured to predicted solar neutrino flux, which is nothing
else but the $\nu_e$ survival probability in Eq.(\ref{Pee}) with all sine
squared terms averaged to 1/2,
\beq
R_e^{Solar} = 1 - 2 \, c^2_\beta \,
              (c^2_\theta s^2_\theta c^2_\beta + s^2_\beta),
\label{ReSolar}
\eeq
yields $R_e^{Solar} = 0.53$ for this particular set of angles.
However, while this solution indicates a significant solar neutrino deficit,
consistent with the current measurements, it is not compatible with other
neutrino experiments, which we discuss in the following sections.

\section{The LSND Experiment}
\begin{figure}[htbp]
\vspace*{13pt}
\begin{center}
\mbox{
\epsfxsize=12cm \epsfbox[25 45 430 545]{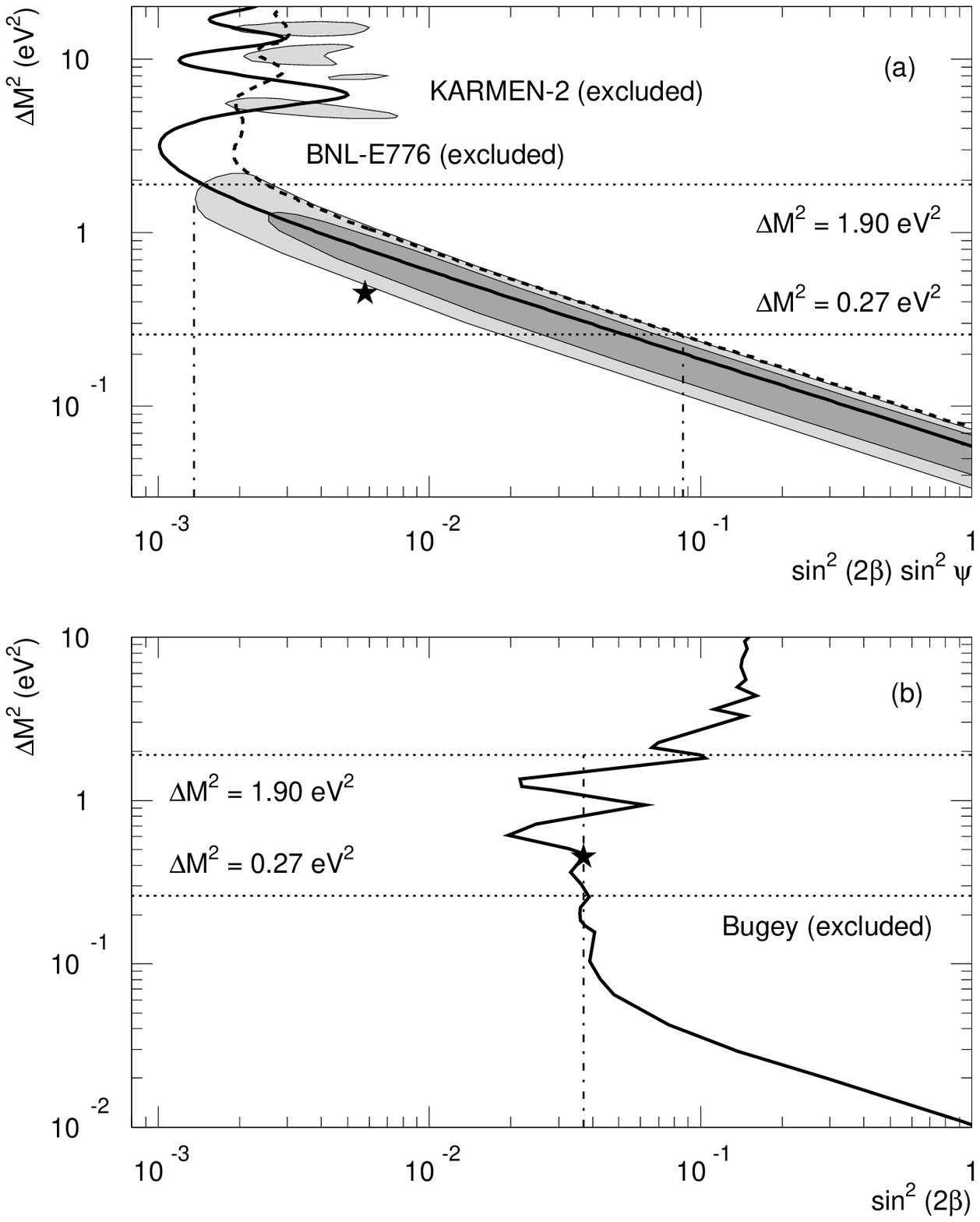}}
\end{center}
\vspace*{13pt}
\fcaption{
(a) Allowed regions in the $(\sin^2 (2\beta) \sin^2 \psi,\Delta M^2)$ space as
obtained by the LSND DAR $\bar\nu_e$ {\footnotesize \it appearance} experiment
(grey-shaded areas).
The excluded regions obtained by the \mbox{BNL-E776} and \mbox{KARMEN-2}
experiments are also shown (thick dashed and thick solid curves, respectively).
The dot-dashed lines correspond to $\sin^2 (2\beta) \sin^2 \psi = 0.0014$ and
$\sin^2 (2\beta) \sin^2 \psi = 0.086$.
(b) Excluded region in the $(\sin^2 (2\beta),\Delta M^2)$ space as obtained by
the Bugey $\bar\nu_e$ {\footnotesize \it disappearance} reactor experiment.
The dot-dashed line corresponds to $\sin^2 (2\beta) = 0.037$.
In both (a) and (b) the dotted lines correspond to the $\Delta M^2$ limits
considered here, while the point favored by this analysis is indicated by the
dark star.
}
\label{Fig-Allowed}
\end{figure}

\noindent
The detector was located at 30 m from the neutrino source, with average
neutrino energies of approximately 40 MeV and 110 MeV, for the decay-at-rest
(DAR) and decay-in-flight (DIF) beams, respectively.
We shall only consider the DAR results, as they are more restrictive and also
more statistically significant than the DIF results.
The $\bar\nu_\mu \to \bar\nu_e$ oscillation probability is given by
Eq.(\ref{Pemu}).
For $\Delta m^2 = {\cal O}(10^{-3})$~eV$^2$ the first sine squared term is
negligible, while the remaining two can be combined by approximating
$\Delta M^2 - \Delta m^2 \approx \Delta M^2$ to yield
\beq
P_{\mu e}^{LSND} = \sin^2 \left( 2 \beta \right) \, \sin^2 \psi \, 
                   \sin^2 \left( 1.27 \, \Delta M^2 \, \frac{L}{E} \right).
\label{PmueLsnd}
\eeq
Comparing this with the two-generation neutrino oscillation expression
\beq
P_{\mu e}^{LSND} = \sin^2 \left( 2 \Theta_{LSND} \right) \,
                   \sin^2 \left( 1.27 \, \Delta M^2 \, \frac{L}{E} \right),
\eeq
in terms of which the LSND event excess has been analyzed, one can effectively
identify $\sin^2 (2 \Theta_{LSND})$ with $\sin^2 (2 \beta) \sin^2 \psi$.
The allowed regions obtained by this experiment from a preliminary analysis of
the entire 1993-1998 DAR data\cite{RexT} is shown in Fig.~\ref{Fig-Allowed}(a).
The excluded regions determined from the \mbox{BNL-E776}\cite{BnlE776} and
\mbox{KARMEN-2}\cite{KlausE} experiments are also shown in the same figure.
Considering these limits, we impose an upper bound for the $\Delta M^2$ mass
scale, $\Delta M^2 < 1.9$~eV$^2$, and a lower bound for the effective mixing,
$\sin^2 (2\beta) \sin^2 \psi > 0.0014$.
Furthermore, following the weak lower bound imposed by the oscillations
averaging requirement, $\Delta M^2 > 0.27$~eV$^2$, implies an upper bound for
the effective mixing, $\sin^2 (2\beta) \sin^2 \psi < 0.086$, as dictated by
the boundary between the allowed region from LSND and the excluded region from
\mbox{BNL-E776}.
These limits determine the allowed region in the $(\beta,\psi)$ mixing angles
space illustrated by the grey-shaded area in Fig.~\ref{Fig-R175}.
The recent \mbox{KARMEN-2} limit would naturally impose a more stringent limit
of $\sin^2 (2\beta) \sin^2 \psi < 0.051$.
However, it turns out that the restrictions imposed by LSND on the angles
$\beta$ and $\psi$ are rather weak as compared to the severe limits imposed by
the Bugey reactor experiment, which we discuss in the next section.

\section{The Reactor Experiments}
\noindent
The survival probability for electron (anti)neutrinos -- as typically
measured by all reactor experiments -- is given by Eq.(\ref{Pee}).
The most restrictive reactor experiment for the relevant range of parameters
considered in this paper is the Bugey experiment.\cite{Bugey}
It was carried out with three detectors at distances of 15 m, 40 m, and 95 m
from the reactor core, with a mean neutrino energy of approximately 3~MeV.
Following the same arguments as in the LSND discussion above, the first sine
squared term in Eq.(\ref{Pee}) is negligible, and the remaining two are
combined to yield
\beq
P_{ee}^{Bugey} = 1 - \sin^2 \left( 2 \beta \right) \,
                     \sin^2 \left( 1.27 \, \Delta M^2 \, \frac{L}{E} \right).
\label{PeeBugey}
\eeq
A direct comparison of Eq.(\ref{PeeBugey}) with Eq.(\ref{PmueLsnd}) shows
explicitly the difference between the disappearance and appearance
probabilities in the three-generations neutrino oscillations formalism,
as we have argued in the previous section.
However, in this particular case, the expression given by Eq.(\ref{PeeBugey})
is fully equivalent with the $\nu_e$ disappearance probability that one would
write in a two-generations mixing model,
\beq
P_{ee}^{Bugey} = 1 - \sin^2 \left( 2 \Theta_{Bugey} \right) \,
                     \sin^2 \left( 1.27 \, \Delta M^2 \, \frac{L}{E} \right).
\eeq
Therefore, one can effectively identify $\sin^2 (2 \Theta_{Bugey})$ with
$\sin^2 (2 \beta)$, or simply $\Theta_{Bugey}$ with $\beta$.
Consequently, the same limits determined for $\Theta_{Bugey}$ apply to the
mixing angle $\beta$ in our formalism, as shown in Fig.~\ref{Fig-Allowed}(b).

Notice that the excluded region obtained by this experiment is not shown on
the same figure with LSND, as one typically tends to do -- see for example
Refs.[18,20].
This is because LSND, \mbox{BNL-E776}, and \mbox{KARMEN-2} are all
\underline{appearance} experiments, whereas Bugey is a
\underline{disappearance} one.
In the two-generations neutrino oscillations formalism, plotting the excluded
regions from a disappearance experiment on the same figure showing the
excluded (or allowed) regions from an appearance experiment is perfectly
legitimate.
However, this is incorrect and misleading in the full three-generations mixing,
where {\it appearance} and {\it disappearance} probabilities are characterized
by different combinations of the mixing angles.

For the $\Delta M^2$ mass scale discussed in this Letter, the Bugey reactor
experiment severely restricts $\sin^2 2\Theta_{Bugey}$ to be $< 0.037$, and
herewith $\beta = \Theta_{Bugey} < 5.5^{\circ}$ -- as indicated by the
hatched area in Fig.~\ref{Fig-R175} -- at the 90\% confidence level.
This limit is obviously much more restrictive than the ones obtained from the
LSND discussion above, limiting the mixing angles $\beta$ and $\psi$ to the two
dark-shaded areas in Fig.~\ref{Fig-R175}.
However, the relevance of the LSND result should not be underestimated.
On one hand, it does not allow for $\beta$ to become arbitrarily small, while
on the other hand -- and more importantly -- it establishes the existence 
of the $\Delta M^2 = {\cal O}(1)$~eV$^2$ mass scale, which is crucial to our
approach.

Before we go on and discuss the mixing matrix determined under these
constraints, we would like to first discuss a general feature that emerges
directly from the narrow intervals to which $\beta$ and $\psi$ are confined.
Both areas of allowed values for $\beta$ and $\psi$ (the upper and lower
dark-shaded areas in Fig.~\ref{Fig-R175}, henceforth referred to as A and B,
respectively) are characterized by a very low mixing angle $\beta < 5.5^\circ$.
This in turn implies that $s_\beta < 0.096$ and herewith, the expression for
the ratio of measured to predicted solar neutrino flux given by
Eq.(\ref{ReSolar}) becomes
\beq
R_e^{Solar} = 1 - \frac{1}{2} \, \sin^2 (2 \theta) + {\cal O}(10^{-2}).
\eeq
Consequently, a mixing angle $\theta$ of approximately $45^\circ$ will
automatically imply a solar neutrino ratio $R_e^{Solar} \simeq 0.5$.
And this is actually exactly what happens.
For values of $\beta$ and $\psi$ restricted to area B in Fig.~\ref{Fig-R175},
the angle $\theta$ must satisfy $20^\circ < \theta < 64^\circ$ for the
predicted ratio of ratios in the 8th $L/E$ bin, as a function of $\theta$, to
agree with the value measured by Super-Kamiokande at the $\pm 1 \sigma$ level,
$R_\mu^{(8)}/R_e^{(8)} = 0.456 \pm 0.069$.
(For mixing angles $\beta$ and $\psi$ in area A of Fig.~\ref{Fig-R175} there
is no $\theta$ solution at the $\pm 2 \sigma$ level.)
This implies that the solar neutrino ratio must be $0.49 < R_e^{Solar} < 0.78$.
However, imposing a stricter agreement with the central value of
$R_\mu^{(8)}/R_e^{(8)} = 0.456$, yields values of $\theta$ of approximately
$35^\circ$ and $50^\circ$, which in turn yield $R_e^{Solar} \simeq 0.56$ and
$R_e^{Solar} \simeq 0.52$, respectively.

\section{The Mixing Matrix and the LSND Mass Scale}
\noindent
Fitting the ratio of ratios and the $\mu$-like event ratios at the two $L/E$
bins, one obtains the following set of mixing angles: $\theta = 35.1^\circ$,
$\beta = 5.5^\circ$, and $\psi = 23.3^\circ$, for an overall flux normalization
factor of 0.87.
Herewith, the mixing matrix reads explicitly
\beq
U = \left(
\begin{array}{rrr}
  0.815 & \,   0.572 & \,\,\,\,\,\, 0.097\\
- 0.559 & \,   0.730 & \,\,\,\,\,\, 0.394\\
  0.155 & \, - 0.375 & \,\,\,\,\,\, 0.914
\end{array}
\right),
\label{MixMatrix}
\eeq
and the ratio of measured to expected solar neutrino flux is predicted to be
$R_e^{Solar} = 0.55$.
Notice that, instead of using the $\mu$-like events, one may very well use the
$e$-like events in determining the mixing angles.
Despite the somewhat larger errors in the $e$-like event ratios, the result
does not differ much from the previous one: $\theta = 32.5^\circ$,
$\beta = 5.5^\circ$, $\psi = 24.6^\circ$, and thus $R_e^{Solar} = 0.58$.

\begin{figure}[htb]
\vspace*{13pt}
\begin{center}
\mbox{
\epsfxsize=8cm \epsfbox[20 35 545 545]{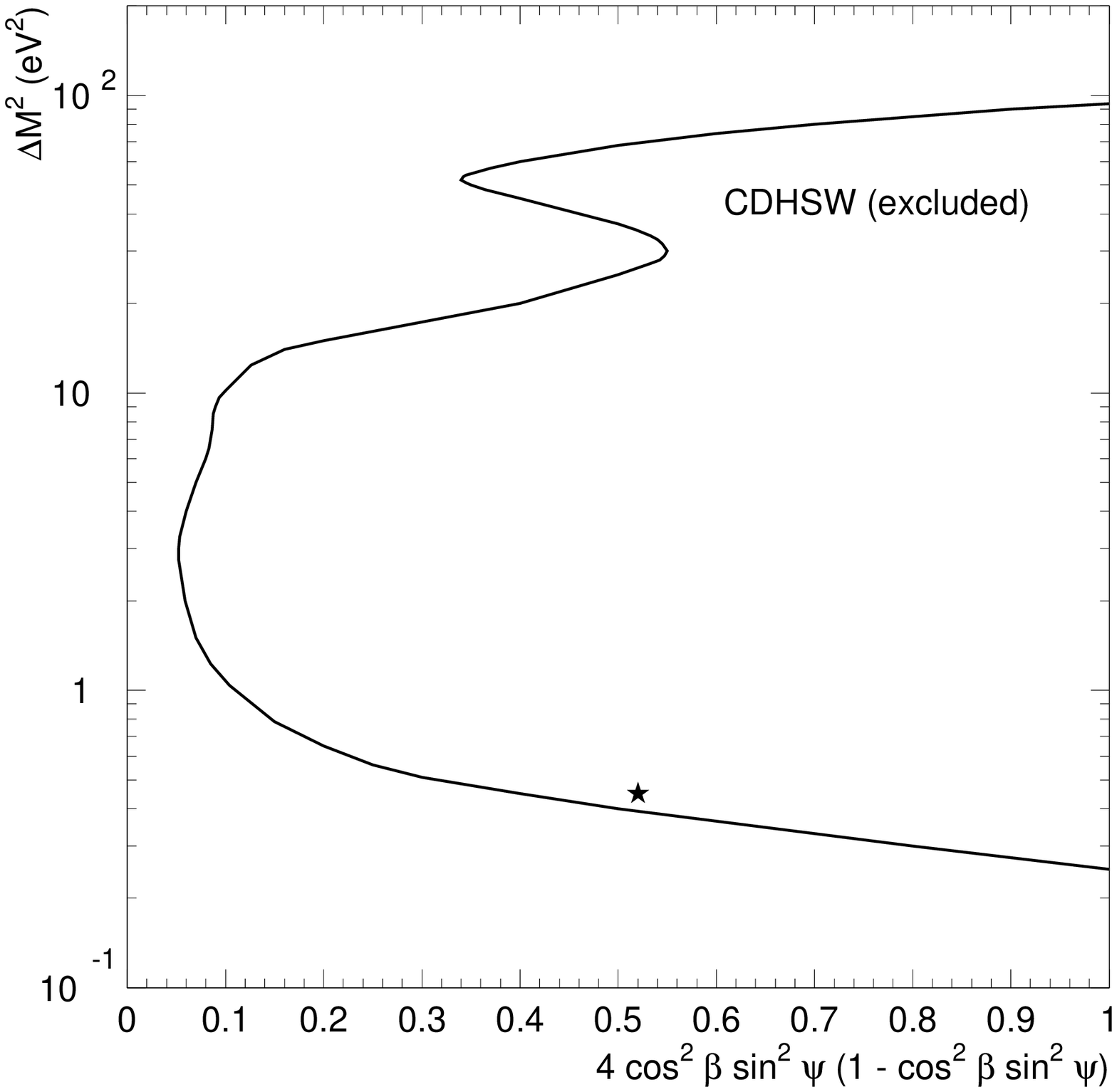}}
\end{center}
\vspace*{13pt}
\fcaption{
Excluded region in the $(4 \cos^2 \beta \sin^2 \psi
(1 - \cos^2 \beta \sin^2 \psi), \Delta M^2)$ space obtained by the CDHSW
$\nu_\mu$ {\footnotesize \it disappearance} experiment, along with the
point favored by this analysis (dark star).
}
\label{Fig-Cdhsw}
\end{figure}

For the mixing angles determined here, $\sin^2 (2 \Theta_{LSND}) = 0.0058$,
and thus $\Delta M^2$ is restricted to 0.5~eV$^2$ $< \Delta M^2 <$ 0.8~eV$^2$,
as one can easily see from the LSND allowed region in
Fig.~\ref{Fig-Allowed}(a), below the area excluded by \mbox{KARMEN-2}.
Thus, the $\Delta M^2$ mass scale range favored by this analysis is well above
the weak limit of approximately 0.27~eV$^2$ imposed by the oscillations
averaging argument.
However, there is yet another mass limit that has to be considered at this
point, namely as imposed by the CDHSW $\nu_\mu$ \underline {disappearance}
experiment\cite{Cdhsw} (illustrated in Fig.~\ref{Fig-Cdhsw}).
This experiment was performed using a neutrino beam of approximately 2~GeV and
two detectors at 130~m and 885~m from the mean neutrino production point.
The $\nu_\mu$ survival probability is given by Eq.(\ref{Pmumu}), which becomes
\beq
P_{\mu\mu}^{CDHSW} = 1 - 4 \cos^2 \beta \, \sin^2 \psi  \,
                    (1 -   \cos^2 \beta \, \sin^2 \psi) \,
                     \sin^2 \left( 1.27 \, \Delta M^2 \, \frac{L}{E} \right),
\eeq
when neglecting the $\Delta m^2$ contribution, as discussed in the previous
two sections.
Comparing this with the two-generations probability
\beq
P_{\mu\mu}^{CDHSW} = 1 - \sin^2 \left( 2 \Theta_{CDHSW} \right) \,
                     \sin^2 \left( 1.27 \, \Delta M^2 \, \frac{L}{E} \right),
\eeq
and using the mixing angles obtained by this analysis ($\beta = 5.5^\circ$ and
$\psi = 23.3^\circ$), one obtains $\sin^2 (2 \Theta_{CDHSW}) = 0.52$.
For this value of the effective mixing, the values of $\Delta M^2$ below
0.4~eV$^2$ are excluded at the 90\% confidence level by this experiment.
Therefore, we conclude that this mass scale must be restricted to
$0.4$~eV$^2$ $< \Delta M^2 <$ 0.5~eV$^2$, and we set it to
$\Delta M^2 \simeq 0.45$~eV$^2$ for the discussion below.
With these parameters, the point favored by this analysis lies at the edge of
the LSND allowed region and also at the edge of the Bugey and CDHSW excluded
regions, as shown in Figs.~\ref{Fig-Allowed}(a), \ref{Fig-Allowed}(b), and
\ref{Fig-Cdhsw}, respectively.

\begin{figure}[htbp]
\vspace*{13pt}
\begin{center}
\mbox{
\epsfxsize=12cm \epsfbox[25 45 430 545]{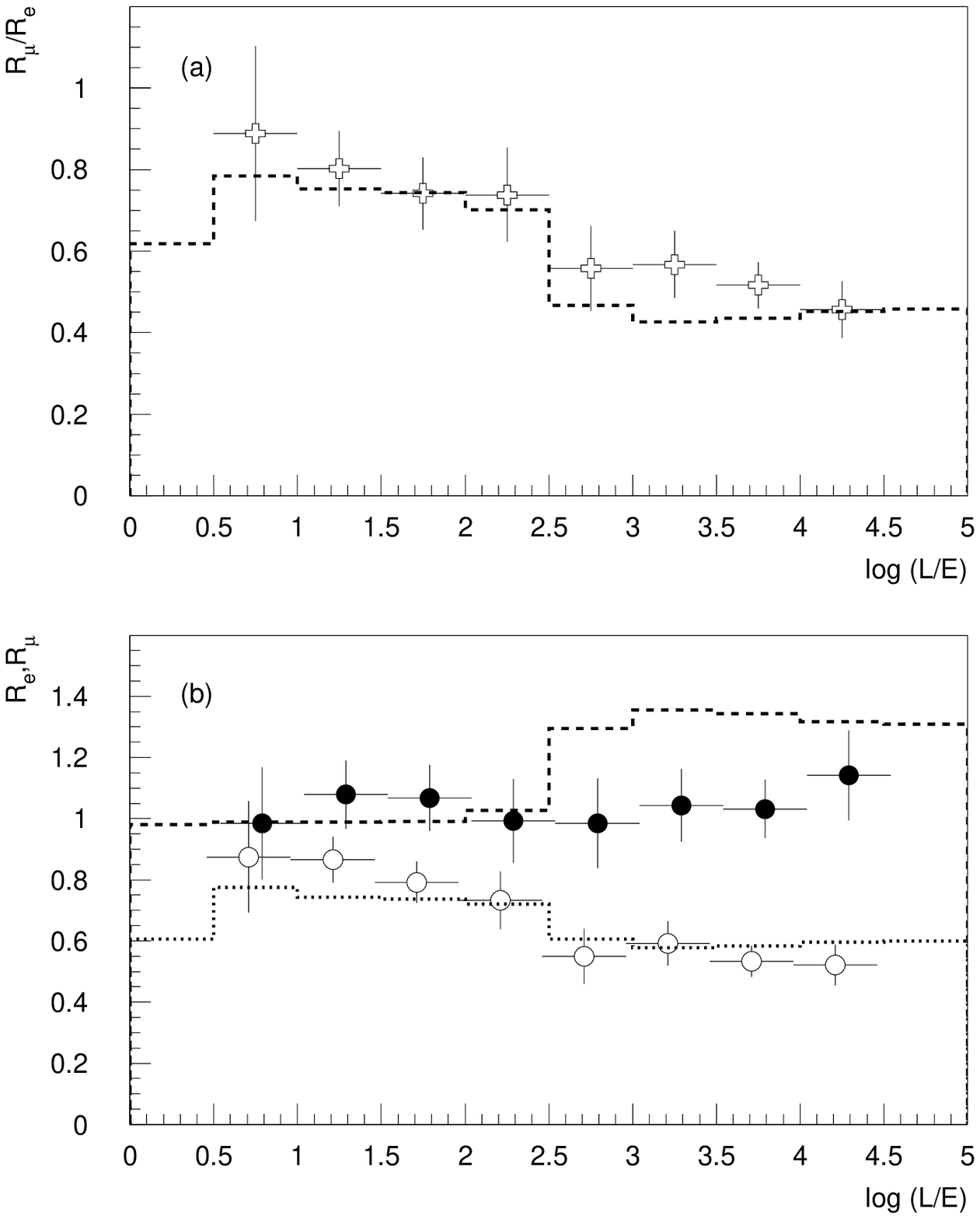}}
\end{center}
\vspace*{13pt}
\fcaption{
(a) The predicted ratio of ratios ($R_\mu/R_e$) versus $\log (L/E)$ --
dashed histogram -- compared to the Super-Kamiokande data (empty crosses with
error bars).
(b) The predicted $\mu$-like (dotted) and $e$-like (dashed) event ratios
($R_\mu$ and $R_e$) versus $\log (L/E)$, compared to the Super-Kamiokande data
(full and empty circles with error bars, respectively) scaled by a factor of
0.87.
Both (a) and (b) have been obtained with $\Delta m^2 = 10^{-3}$~eV$^2$ and
$\Delta M^2 = 0.45$~eV$^2$.
}
\label{Fig-RPredicted}
\end{figure}

With the mixing matrix fully determined and with
$\Delta m^2 = {\cal O}10^{-3}$~eV$^2$
and
$\Delta M^2 \simeq 0.45$~eV$^2$,
we can also investigate the atmospheric neutrino ratio of ratios predicted by
this analysis as compared to the Super-Kamiokande data.
This is illustrated in Fig.~\ref{Fig-RPredicted}(a) for
$\Delta m^2 = 10^{-3}$~eV$^2$, as we shall make no attempt in this Letter to
fit this mass scale.
Our predicted $\mu$-like and $e$-like event ratios are shown in
Fig.~\ref{Fig-RPredicted}(b), together with the Super-Kamiokande data
rescaled by an overall factor of 0.87.
In both cases the agreement between our prediction and the experimental data
is reasonably good.
No smearing of $L/E$ has been included in our simulations.

\section{Conclusions}
\noindent
We have shown that the $L/E$ dependence of the Super-Kamiokande atmospheric
neutrino data fully determines the three-generations neutrino mixing matrix,
as given by Eq.(\ref{MixMatrix}).
The only underlying assumption is that the mass scales involved are those
determined by the atmospheric neutrino data,
$\Delta m^2 = {\cal O}(10^{-3})$~eV$^2$,
and by the LSND experiment,
$\Delta M^2 = {\cal O}(1)$~eV$^2$.
Consequently, the ratio of measured to predicted solar neutrino flux is
determined to be $R_e^{Solar} = 0.55$, independent of energy.
While no attempt has been made here to determine a more precise value for
$\Delta m^2$, we believe that $\Delta M^2 \simeq 0.45$~eV$^2$.
This, together with the effective $\sin^2 (2 \Theta_{LSND}) = 0.0058$
determined here, lies just outside the sensitivity of the \mbox{KARMEN-2}
experiment, to be reached at the end of 2001.
However, the MiniBooNE experiment at FermiLab should see a significant number
of $\nu_\mu \to \nu_e$ events at this mixing: approximately 345 (525) per year
for $\Delta M^2 = 0.4 \,\, (0.5)$~eV$^2$.
The long-baseline $\nu_\mu \to \nu_\tau$ appearance experiments (K2K, MINOS),
which are probing the $\Delta m^2$ mass sale,
should also see significant excess events for this particular neutrino mixing
pattern.
Higher statistics in the Super-Kamiokande atmospheric neutrino data and a more
detailed analysis -- which may include deviations of $r = N_\mu/N_e$ from the
naive value of $r = 2$ due to geo-magnetic effects and $L/E$ variations -- 
should further restrict the mixing angles in this neutrino oscillations model,
which involves the $\Delta m^2 = {\cal O}(10^{-3})$~eV$^2$ and
$\Delta M^2= {\cal O}(1)$~eV$^2$ mass scales.
However, should the LSND result prove to be wrong, it might be just as simple
as: atmospheric deficit $\equiv \nu_\mu \to \nu_\tau$, and solar deficit
$\equiv \nu_e \to \nu_x$.
On the other hand, should the LSND result stand and should the solar neutrino
deficit show a strong energy dependence, more complex neutrino models have to
be invoked.
Either way, if neutrino oscillations are firmly established, this will open
exciting avenues beyond the physics of the Standard Model.

\nonumsection{References}


\noindent
\begin{thebibliography}{000}
\bibitem{Pontecorvo} B.\ Pontecorvo,
                     {\bibit Zh.\ Eksp.\ Teor.\ Fiz.} {\bibbf 33}, 549 (1957);
                     {\bibit Sov.\ Phys.\ JETP} {\bibbf 6}, 429 (1958).
\bibitem{Homestake}  B.\ T.\ Cleveland {\bibit et al.},
                     {\bibit Nucl.\ Phys.\ B} {\bibbf 38}
                     {\bibit (Proc.\ Suppl.)}, 47 (1995).
\bibitem{Kamiokande} Y.\ Fukuda {\bibit et al.},
                     {\bibit Phys.\ Rev.\ Lett.} {\bibbf 77}, 1683 (1996).
\bibitem{Sage}       J.\ N.\ Abdurashitov {\bibit et al.},
                     {\bibit Phys.\ Lett.\ B} {\bibbf 328}, 234 (1994).
\bibitem{Gallex}     W.\ Hampel {\bibit et al.},
                     {\bibit Phys.\ Lett.\ B} {\bibbf 388}, 364 (1996).
\bibitem{SkSol}      Y.\ Fukuda {\bibit et al.}
                     (Super-Kamiokande Collaboration),
                     hep-ex/9812011.
\bibitem{Hirata}     K.\ S.\ Hirata {\bibit et al.},
                     {\bibit Phys.\ Lett.\ B} {\bibbf 280}, 146 (1992).
\bibitem{Becker}     R.\ Becker-Szendy {\bibit et al.},
                     {\bibit Phys.\ Rev.\ D} {\bibbf 46}, 3720 (1992).
\bibitem{SkAtm}      Y.\ Fukuda {\bibit et al.}
                     (Super-Kamiokande Collaboration),
                     {\bibit Phys.\ Rev.\ Lett.} {\bibbf 81}, 1562 (1998).
\bibitem{LsndDar935} C.\ Athanassopoulos {\bibit et al.}
                     (LSND Collaboration),
                     {\bibit Phys.\ Rev.\ C} {\bibbf 54}, 2685 (1996).
\bibitem{LsndDif935} C.\ Athanassopoulos {\bibit et al.}
                     (LSND Collaboration),
                     {\bibit Phys.\ Rev.\ C} {\bibbf 58}, 2489 (1998).
\bibitem{Thun}       R.\ P.\ Thun and S.\ McKee,
                     {\bibit Phys.\ Lett.\ B} {\bibbf 439}, 123 (1998).
\bibitem{Barenboim}  G.\ Barenboim and F.\ Scheck,
                     {\bibit Phys.\ Lett.\ B} {\bibbf 440}, 332 (1998).
\bibitem{Teshima}    T.\ Teshima and T.\ Sakai, hep-ph/9805386, hep-ph/9901219.
\bibitem{Ohlsson}    T.\ Ohlsson and H.\ Snellman, hep-ph/9903252.
\bibitem{Maiani}     L.\ Maiani,
                     {\bibit Phys.\ Lett.\ B} {\bibbf 62}, 183 (1976).
\bibitem{DvaTg}      D.\ V.\ Ahluwalia and T.\ Goldman,
                     {\bibit Phys.\ Rev.\ D} {\bibbf 56}, 1698 (1997).
\bibitem{RexT}       R.\ Tayloe, talk presented at the {\bibit Lake Louise
                     Winter Institute}, Lake Louise, Canada, 14-20 February,
                     1999.
\bibitem{BnlE776}    L.\ Borodovsky {\bibit et al.},
                     {\bibit Phys.\ Rev.\ Lett.} {\bibbf 68}, 274 (1992).
\bibitem{KlausE}     K.\ Eitel, talk presented at the {\bibit Lake Louise
                     Winter Institute}, Lake Louise, Canada, 14-20 February,
                     1999.
\bibitem{Bugey}      B.\ Achkar {\bibit et al.},
                     {\bibit Nucl.\ Phys.\ B} {\bibbf 434}, 503 (1995).
\bibitem{Cdhsw}      F.\ Dydak {\bibit et al.},
                     {\bibit Phys.\ Lett.\ B} {\bibbf 134}, 281 (1984).
\end{thebibliography}
\end{document}